\documentclass{PoS}

\usepackage{amsmath}
\usepackage{booktabs}
\usepackage{cite}
\usepackage[english]{babel}

\def\beq{\begin{equation}}  
\def\eeq{\end{equation}}
\def\({\left(}
\def\){\right)}
\def\[{\left[}
\def\]{\right]}

\title{Improved description of the HERA data with a new simple PDF parametrization}

\ShortTitle{A new simple PDF parametrization}

\author{\speaker{Francesco Giuli}\\
        University of Rome Tor Vergata and INFN, Sezione di Roma 2,\\ Via della Ricerca Scientifica~1, 00133 Roma, Italy\\
        E-mail: \email{francesco.giuli@roma2.infn.it}}

\author{Marco Bonvini\\
        INFN, Sezione di Roma 1,\\ Piazzale Aldo Moro~5, 00185 Roma, Italy\\
        E-mail: \email{marco.bonvini@roma1.infn.it}}

\abstract{A new parametrization for the parton distribution functions with a higher  flexibility in the small-$x$ region is presented. It has been implemented in the \texttt{xFitter} open-source PDF fitting tool, and compared to the default \texttt{xFitter} parametrization, used for the determination of the HERAPDF set. It has been found that the combined inclusive HERA~I+II data can be described using NNLO theory with a significantly higher quality than HERAPDF2.0: the $\chi^2$ is reduced by more than 60 units, having used only four more parameters. Our result highlights a significant parametrization bias in the default \texttt{xFitter} parametrization at small $x$, which would lead to even more dramatic effects when used for higher energy colliders, where the small-$x$ region is more relevant. We also find that the inclusion of small-$x$ resummation leads to a further reduction by approximately 30 extra units in $\chi^2$. In this contribution, we review the results of the recent paper "A new simple PDF parametrization: improved description of the HERA data" (arXiv:1902.11125).}

\FullConference{XXVII International Workshop on Deep-Inelastic Scattering and Related Subjects - DIS2019\\
		8-12 April, 2019\\
		Torino, Italy}

\begin{document}

\section{The new proposed parametrization}
Among many others, parton distribution functions (PDFs) represent a fundamental aspect of perturbative QCD (pQCD) in presence of incoming protons. These object describe the longitudinal momentum fraction \textit{x} of partons within the proton.
Currently the most accurate and reliable way to determine PDFs is through fits to data; thus the resulting fitted distributions depend on various aspects of this procedure e.g. the perturbative order of partonic cross sections or DGLAP splitting functions, the way heavy quarks are treated, $\chi^2$ definition, minimizations methods, the choice of the PDF  parametrisation at the initial scale $Q_0^{2}$, etc.
Here we review the results of Ref.~\cite{Bonvini:2019wxf}, where a new flexible and simple parametrization has been proposed and successfully used to determine PDFs.

Starting from the default parametrization used in \texttt{xFitter}~\cite{Alekhin:2014irh,Bertone:2017tig} (which is the one used for the HERAPDF set), namely
\begin{equation}
\label{eq:OldPar}
xf(x,\mu_0^2) = A\,x^B (1-x)^C \Big[1+Dx+Ex^2\Big] -A'\,x^{B'} (1-x)^{C'},
\end{equation}
we propose a new extension, designed to add more flexibility in the small-$x$ region,
while keeping the number of fitted parameters small.
A polynomial in $\log x$, which gives flexibility in the low-$x$ region,
is added on top of the polynomial in $x$, which gives flexibility in the high-$x$ region.
These two polynomials can be combined considering a multiplicative option
\begin{equation}
\label{eq:NewParMult}
xf(x,\mu_0^2) = A\,x^B (1-x)^C \Big[1+Dx+Ex^2\Big] \Big[1+F\log x+G\log^2x+H\log^3x\Big]
\end{equation}
or an additive option
\begin{equation}
\label{eq:NewParAdd}
xf(x,\mu_0^2) = A\,x^B (1-x)^C \Big[1+Dx+Ex^2+F\log x+G\log^2x+H\log^3x\Big].
\end{equation}
These two options have been tested and it has been found that the additive parametrization results
in smoother shapes and smaller $\chi^2$ in the fit.
So, the actual parametrization used in our fits to the inclusive HERA data is
\begin{subequations}\label{eq:NewPar}
\begin{align}
  xg(x,\mu_0^2) &= A_g\,x^{B_g} (1-x)^{C_g} \Big[1+F_g\log x+G_g\log^2x\Big] \\
  xu_v(x,\mu_0^2) &= A_{u_v}\,x^{B_{u_v}} (1-x)^{C_{u_v}} \Big[1+E_{u_v}x^2+F_{u_v}\log x+G_{u_v}\log^2x\Big] \\
  xd_v(x,\mu_0^2) &= A_{d_v}\,x^{B_{d_v}} (1-x)^{C_{d_v}} \\
  x\bar u(x,\mu_0^2) &= A_{\bar u}\,x^{B_{\bar u}} (1-x)^{C_{\bar u}} \Big[1+D_{\bar u}x+F_{\bar u}\log x\Big] \\
  x\bar d(x,\mu_0^2) &= A_{\bar d}\,x^{B_{\bar d}} (1-x)^{C_{\bar d}} \Big[1+D_{\bar d}x+F_{\bar d}\log x\Big].
\end{align}
\end{subequations}
This new parametrization depends on 18 free parameters to be fitted at the starting scale, which is to be compared with the HERAPDF2.0 parametrization, which depends on 14 free parameters.

\section{PDF determination at NNLO}
In order to directly compare our fit results with HERAPDF2.0, we use the same definition of the $\chi^2$, namely~\cite{Abramowicz:2015mha}
\begin{equation}
\label{eq:chi2}
  \chi^2 = \sum_i\frac{\[ D_i -T_i\(1-\sum_j\gamma_{ij} b_j\) \]^2}
  {\delta_{i,{\rm uncor}}^2 T_i^2+\delta_{i,{\rm stat}}^2D_i T_i}
  + \sum_j b_j^2
  + \sum_i \log \frac{\delta_{i,{\rm uncor}}^2 T_i^2+\delta_{i,{\rm stat}}^2D_i T_i}
  {\delta_{i,{\rm uncor}}^2 D_i^2+\delta_{i,{\rm stat}}^2D_i^2},
\end{equation}
where the measured data are represented by $D_i$, the corresponding theoretical prediction by $T_i$, $\delta_{i,\rm uncor}$ and $\delta_{i,\rm stat}$ represent the uncorrelated systematic and the statistical uncertainties on the measured data respectively, $\gamma_{ij}$ describe the correlated systematics which are accounted for using the nuisance parameters $b_j$. The sums over $i$ extend over all data points, while the sum over $j$ runs over the various sources of correlated systematics.

\begin{table}
\small
\centering
\begin{tabular}{lll}
  Differences in the fit setup & Old setup, same as~\cite{Abramowicz:2015mha} & New setup, same as~\cite{Abdolmaleki:2018jln}  \\
  \midrule
  heavy flavour scheme & TR & FONLL \\
  initial scale $\mu_0$ & $1.38$~GeV & 1.6~GeV \\
  charm matching scale $\mu_c$ & $m_c$ & $1.12m_c$ \\
  charm mass $m_c$ & 1.43~GeV & 1.46~GeV
\end{tabular}
\caption{Summary of the differences in the theoretical setup between
  the fit of HERAPDF2.0~\cite{Abramowicz:2015mha}
  and the new fits presented in this and in the following sections
  (which is the same of Ref.~\cite{{Abdolmaleki:2018jln}}).}
\label{tab:diff}
\end{table}
\begin{table}
\small
\centering
\begin{tabular}{lcc}
  Contribution to $\chi^2$ & Old parametrization~\cite{Abdolmaleki:2018jln} &  New parametrization  \\
  \midrule
  subset NC $e^+$ 920    $\tilde\chi^2/\rm{n.d.p.}$   & $451/377$   & $406/377$   \\
  subset NC $e^+$ 820    $\tilde\chi^2/\rm{n.d.p.}$   & $ 68/ 70$   & $ 74/ 70$   \\
  subset NC $e^+$ 575    $\tilde\chi^2/\rm{n.d.p.}$   & $220/254$   & $222/254$   \\
  subset NC $e^+$ 460    $\tilde\chi^2/\rm{n.d.p.}$   & $218/204$   & $225/204$   \\
  subset NC $e^-$        $\tilde\chi^2/\rm{n.d.p.}$   & $215/159$   & $217/159$   \\
  subset CC $e^+$        $\tilde\chi^2/\rm{n.d.p.}$   & $ 44/ 39$   & $ 37/ 39$   \\
  subset CC $e^-$        $\tilde\chi^2/\rm{n.d.p.}$   & $ 57/ 42$   & $ 50/ 42$   \\
  correlation term + log term     & $100+15$ & $79+2$ \\
  \boldmath Total $\chi^2/\rm{d.o.f.}$  &\boldmath $1388/1131$ &\boldmath $1312/1127$ \\
\end{tabular}
\caption{Total $\chi^2$ per degrees of freedom (d.o.f.) and the partial $\tilde\chi^2$  per number of data points (n.d.p.) of each subset of the inclusive HERA dataset, for HERAPDF2.0 and our fit obtained with the parametrization Eq.~\eqref{eq:NewPar}. The second and third terms of Eq.~\eqref{eq:chi2}, denoted correlation and log terms respectively, are also shown. The FONLL scheme is used, having raised $\mu_c/m_c=1.12$, $\mu_0=1.6$~GeV and $m_c=1.46$~GeV, namely the setting of Ref.~\cite{Abdolmaleki:2018jln}.}
\label{tab:chi2TR}
\end{table}
Instead of the ``optimized'' version~\cite{Thorne:2012az} of the Thorne-Roberts scheme~\cite{Thorne:1997ga,Thorne:2006qt} used in HERAPDF2.0~\cite{Abramowicz:2015mha}, the FONLL scheme~\cite{Forte:2010ta} is used as heavy quark mass scheme. The differences between the two fit setups are summarized in Tab.~\ref{tab:diff}.
The results of the fit in terms of $\chi^2$, switching from the old parametrization to the new one in FONLL scheme, are presented in Tab.~\ref{tab:chi2TR}. A significant reduction of 76 units of the total $\chi^2$ is observed, which is much larger than the increase of 4 units in the number of parameters.

\begin{figure}[t]
  \centering
  \includegraphics[width=0.328\textwidth,page=1]{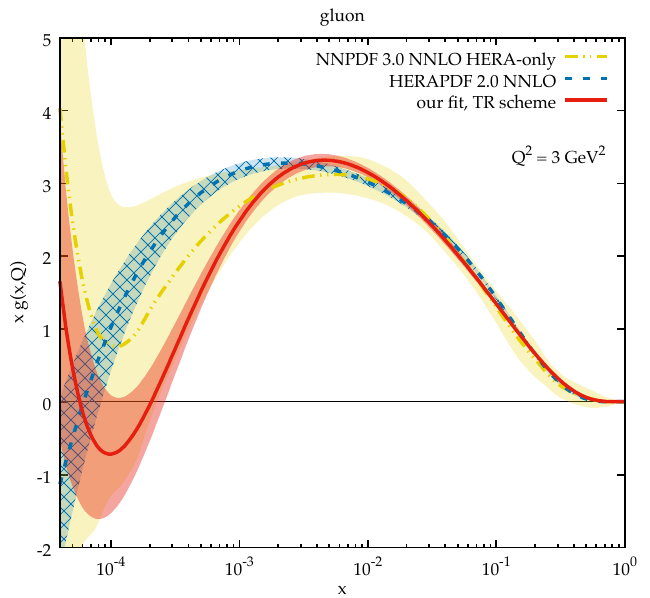}
  \includegraphics[width=0.328\textwidth,page=5]{NNPDF_HERAPDF_comparison.pdf}
  \includegraphics[width=0.328\textwidth,page=6]{NNPDF_HERAPDF_comparison.pdf}
  \caption{Comparison of our fit (solid red) with HERAPDF2.0 (dashed blue) and NNPDF3.0 HERA-only (dot-dot-dashed yellow) for the gluon, $u_v$ and $d_v$ PDFs. The uncertainty shown is only the ``experimental'' one, namely the one coming from the uncertainty on the parameters determined from the fit. For NNPDF, this uncertainty actually covers other kinds of uncertainties, such as those coming from parametrization bias.}
\label{fig:TR}
\end{figure}
Moving to the PDF comparison, the gluon, $u_v$ and $d_v$ distributions at the scale $Q^{2} =$ 3 GeV$^{2}$ are shown in Fig.~\ref{fig:TR}, where our fit results are plotted along with the HERAPDF2.0.
It can be noticed that the shape is generally smoother for HERAPDF2.0, while a richer structure in the medium- and small-$x$ region is present in our PDFs. The comparison between our PDFs and a NNPDF3.0 set obtained fitting only HERA data~\cite{Ball:2014uwa} is also shown in Fig.~\ref{fig:TR}. This choice has been made because NNPDF has the msot flexible parametrization available on the market. It is noticeable that our PDFs lie inside the NNPDF uncertainty bands in most regions of $x$, while the HERAPDF2.0 PDFs lie outside in many more cases. Furthermore, we observe that the gluon shape predicted by NNPDF is very similar to ours and instead quite different from HERAPDF2.0.

\begin{figure}[t]
  \centering
  \includegraphics[width=0.76\textwidth]{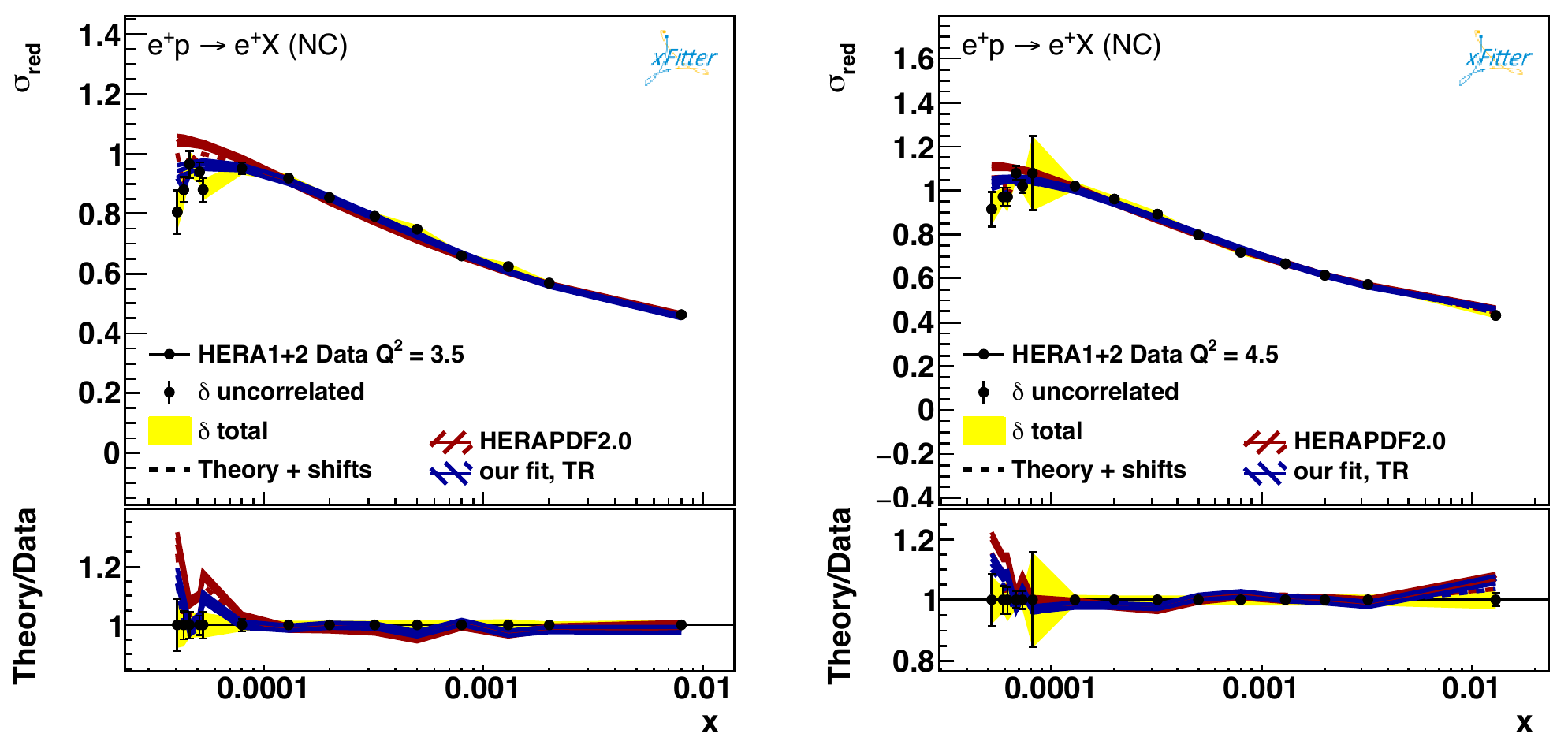}
  \caption{An example of comparison of the theoretical predictions from our     and HERAPDF2.0 with data: low-$Q^2$ neutral-current reduced cross section from the neutral-current $E_p=920$~GeV dataset.}
  \label{fig:datacomp}
\end{figure}
The comparison of HERA data with theoretical predictions using both our fit and HERAPDF2.0 has been inspected in detail. It can been seen in Tab.~\ref{tab:chi2TR} that the agreement is at the same level, apart from the low-$x$ and low-$Q^{2}$ data (contained in the first dataset of the list), where a clear improvement is manifest. Fig.~\ref{fig:datacomp} shows the two lowest $Q^{2}$ bins included in the fit, namely at $Q^{2} =$ 3.5 GeV$^2$ and $Q^{2} =$ 4.5 GeV$^2$, which contain the data at lowest $x$; here, the flexibility of our parametrization in the small-$x$ regime allows to better describe this region.

\section{Reducing correlations for a stabler fit}
A strong correlation between the parameters of the fit may lead to instabilities. In our parametri\-zation, the parameters governing the small-$x$ region ($B,F,G,H$) turned out to have significant correlations.
In order to reduce such correlation, it is useful to redefine the parameters
such that they each multiply a function whose contribution is dominant in a restricted region.
Bernstein polynomials provide an easy way to achieve this goal. For instance, a generic polynomial of degree $n$
in $x$ in the range $0<x<1$ can be conveniently expressed as a linear combination of the Bernstein basis polynomials
\beq
{n\choose k}
(1-x)^k x^{n-k}, \qquad k=0,\ldots,n,
\eeq
each of which is peaked in a different region of $x$.
The variable $x$ can also be replaced by a function of $x$ which still ranges from 0 to 1.
For instance in Ref.~\cite{Dulat:2015mca} this basis was used in CT fits, but replacing $x$ with $\sqrt{x}$.
In our case, the most obvious choice is to use $\log\frac1x$, which however ranges from 0 to infinity.
To circumvent this difficulty, we simply consider a limited range in $x$ in which we reparametrize our polynomial in
$\log\frac1x$ in terms of Bernstein polynomials. Since the data only extend to a small but finite value of $x$,
we consider the range $x_0<x<1$, with $x_0\sim10^{-4}$.
Therefore, we can use as a basis for our parametrization the polynomials
\beq
{n\choose k}
(1-y(x))^k y(x)^{n-k}, \qquad k=0,\ldots,n,
\qquad
{\rm with}\quad y(x) =\frac{\log\frac1x}{\log\frac1{x_0}}.
\eeq
In our specific case, we actually mix a polynomial in $x$ and in $\log\frac1x$, Eq.~\eqref{eq:NewParAdd}.
These two variables, or better $x$ and $y(x)$, tend to zero in opposite limits, and therefore describe opposite regions.
The best option to separate off the two regions described by these two polynomials is to consider two different Bernstein polynomials, one in $x$ and one in $y(x)$, suppressing each with the $k=n$ contribution of the other.
However, this option does no longer correspond to the polynomial we used in our fits, due to the presence of contributions $x^a\log^b\frac1x$ with both $a>0$ and $b>0$, which are absent in Eq.~\eqref{eq:NewParAdd}.
Therefore, we propose a simpler choice, in which the $x$ polynomial is treated as a ``correction'' to the $\log\frac1x$ polynomial. Our most generic parametrization Eq.~\eqref{eq:NewParAdd} then becomes
\begin{align}
xf(x,\mu_0^2)
  &= A\,x^B (1-x)^C \Big[1+F\log x+G\log^2x+H\log^3x+Dx+Ex^2\Big] \\
  &= A\,x^B (1-x)^C \Big[(1-y)^3+3F'y(1-y)^2+3G'y^2(1-y)+H'y^3+2D'x(1-x)+E'x^2\Big], \nonumber
\end{align}
where the new ``primed'' parameters should be much less correlated among each other,
thereby leading to a stabler minimization procedure.
Simpler versions with less coefficients can be constructed in similar ways.
For instance, when the $\log^3x$ term is not used, as in our default parametrization Eq.~\eqref{eq:NewPar}, we simply have
\begin{align}
xf(x,\mu_0^2)
  &= A\,x^B (1-x)^C \Big[1+F\log x+G\log^2x+Dx+Ex^2\Big] \nonumber\\
  &= A\,x^B (1-x)^C \Big[(1-y)^2+2F''y(1-y)+G''y^2+2D'x(1-x)+E'x^2\Big].
\end{align}
Similarly, one can switch off either the $D$ or the $E$ term, which leaves the other unmodified.
We plan to test this new form of the parametrization once the new version of \texttt{xFitter} will be released.

\section{\boldmath PDF determination with small-$x$ resummation}
It has been observed that much of the improvement in the $\chi^2$ when using our new parametrization comes from a better description of the low-$x$ low-$Q^2$ data which are also responsible for the success of small-$x$ resummation~\cite{Ball:2017otu,Abdolmaleki:2018jln}.
Moreover, the reduction in $\chi^{2}$ obtained using our new parametrization is of the same size as the one obtained when including small-$x$ resummation effect in theory~\cite{Abdolmaleki:2018jln}.
In order to understand the interplay between the inclusion of small-$x$ resummation and the use of our new parametrization, PDF fits including small-$x$ resummation with our new parametrization have been performed.

\begin{table}
\small
\centering
\begin{tabular}{lccc}
  Contribution to $\chi^2$ & \texttt{HELL3.0} (NLL) &  \texttt{HELL3.0} (LL$'$) & \texttt{HELL2.0} (LL$'$)  \\
  \midrule
  subset NC $e^+$ 920    $\tilde\chi^2/\rm{n.d.p.}$   & $402/377$   & $403/377$   & $403/377$ \\
  subset NC $e^+$ 820    $\tilde\chi^2/\rm{n.d.p.}$   & $ 70/ 70$   & $ 69/ 70$   & $ 69/ 70$ \\
  subset NC $e^+$ 575    $\tilde\chi^2/\rm{n.d.p.}$   & $219/254$   & $219/254$   & $218/254$ \\
  subset NC $e^+$ 460    $\tilde\chi^2/\rm{n.d.p.}$   & $223/204$   & $224/204$   & $224/204$ \\
  subset NC $e^-$        $\tilde\chi^2/\rm{n.d.p.}$   & $219/159$   & $220/159$   & $220/159$ \\
  subset CC $e^+$        $\tilde\chi^2/\rm{n.d.p.}$   & $ 38/ 39$   & $ 38/ 39$   & $ 38/ 39$ \\
  subset CC $e^-$        $\tilde\chi^2/\rm{n.d.p.}$   & $ 49/ 42$   & $ 49/ 42$   & $ 49/ 42$ \\
  correlation term + log term                         & $73-7$      & $72-11$     & $72-10$ \\
  \boldmath Total $\chi^2/\rm{d.o.f.}$  &\boldmath $1284/1127$ &\boldmath $1283/1127$ &\boldmath $1283/1127$ \\
\end{tabular}
\caption{Same as Tab.~\ref{tab:chi2TR}, for three variants of the resummed NNLO+NLL$x$ fit using our new parametrization.}
\label{tab:chi2res}
\end{table}
The inclusion of small-$x$ resummation corrections is achieved using the \texttt{HELL} code~\cite{Bonvini:2016wki,Bonvini:2017ogt,Bonvini:2018xvt,Bonvini:2018iwt}. Here, three different variants of the resummed NNLO+NLL$x$ fit have been performed; these fits differ from each other in the treatment of subleading logarithmic contributions. Tab.~\ref{tab:chi2res} present the various $\chi^2$ contributions; it is immediately noticeable that the three fits are of the same quality and in all the cases the $\chi^2$ reduction with respect to the NNLO fit (third column in Tab.~\ref{tab:chi2TR}) is about 30 units less.

\begin{figure}[t]
  \centering
  \includegraphics[width=0.388\textwidth,page=1]{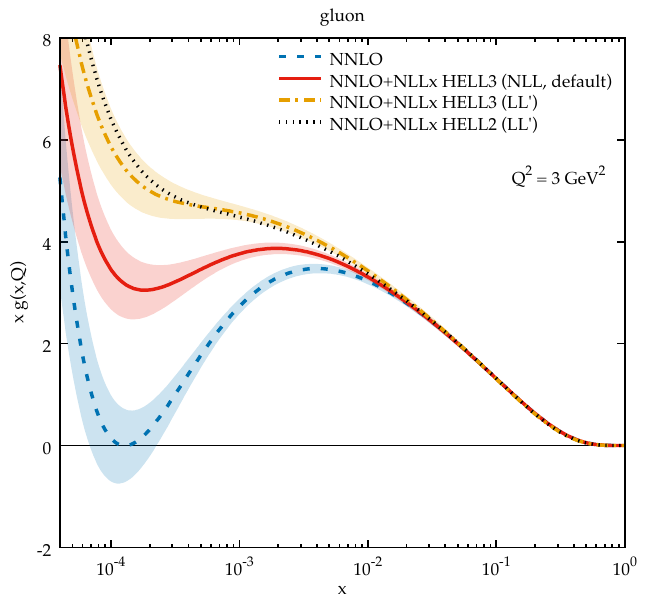}
  \includegraphics[width=0.388\textwidth,page=7]{HELL_comparison.pdf}
  \caption{Comparison of PDFs obtained including small-$x$ resummation from different versions and variants of the \texttt{HELL} code. The band represents only the fit uncertainty. The NNLO fit is also shown for reference.}
  \label{fig:HELL}
\end{figure}
Moving to the PDFs comparison, the gluon PDF is shown at $Q^2 =$ 3 GeV$^{2}$ and in form of ratios at $Q^2 =$ 10$^4$ GeV$^2$ in Fig.~\ref{fig:HELL}.
We conclude that even though subleading logarithmic contributions may change the size of the effect of resummation on the PDFs, the resummed version of the gluon and the quark-singlet PDFs are always significantly larger at small-$x$ than at NNLO.

\end{document}